\documentclass[prc,twocolumn,preprintnumbers,amsmath,amssymb,letter,floatfix]{revtex4}
\usepackage{sublabel}
\usepackage{graphicx}
\usepackage{epsf}
\usepackage{wrapfig}
\usepackage{epsfig}
\usepackage{wasysym}
\def\Vec#1{\mbox{\boldmath $#1$}}

\def\beq{\begin{equation}}
\def\eeq{\end{equation}}

\def\beqy{\begin{eqnarray}}
\def\eeqy{\end{eqnarray}}

\begin{document}
\preprint{CERN-PH-TH/2010-036}
\date{today}
\title{Proton Decay: Improving the sensitivity through nuclear dynamics?}
\vskip 0.5cm
\author{M. Alvioli}
\affiliation{104 Davey Lab, The Pennsylvania State University,
  University Park, PA 16803, USA}
\author{O. Benhar}
\affiliation{INFN, Sezione Roma 1, Dipartimento di Fisica,
Universit\`a "La Sapienza", I-00185 Roma, Italy}
\author{M. Ericson}
\affiliation{Universit\'e de Lyon, Univ.  Lyon 1,
  CNRS/IN2P3, IPN Lyon, F-69622 Villeurbanne Cedex}
\affiliation{Theory Group, Physics Department, CERN, CH-1211, Gen\`eve 23, Switzerland}
\author{M. Strikman}
\affiliation{104 Davey Lab, The Pennsylvania State University,
  University Park, PA 16803, USA{ }}
\begin{abstract}
The kinematics of the decay of a bound proton is governed by the proton spectral function.
We evaluate this quantity in $^{16}O$ using the information from nuclear physics experiments.
It also includes  a correlated part. The reliability of this evaluation is sufficient to open
the possibility of correlated cuts in the missing mass and momentum variables in order to
identify the decay events from the bound protons with a possible increase of the signal to
noise ratio.
\end{abstract}
%\author{
%M. \textsc{Alvioli}$^{1}$\footnote{ e-mail address: alvioli@pg.infn.it},
%O. \textsc{Benhar}$^{2}$\footnote{ e-mail address: Omar.Benhar@roma1.infn.it},
%M. \textsc{Ericson} $^{3,4}$\footnote{ e-mail address: magda.ericson@cern.ch},
%and M. \textsc{Strikman} $^{1}$\footnote{ e-mail address: strikman@phys.psu.edu}
%}
%------------------------------------------------------------------
%------------------------------------------------------------------
\maketitle
%------------------------------------------------------------------

\section{Introduction}

Proton decay is an important window for theories beyond the standard model. Several decay
channels have been experimentally explored, leading to lower limits for the corresponding
proton lifetimes. One of them is the pionic decay mode into a neutral pion and a positron.
The signature of a decay process is that the sum of the four momenta of the decay products
should reconstitute the proton four-momentum. For a free proton at rest this means a vanishing
total three-momentum and a missing mass equal to the free proton mass.
The pionic decay experiment at Kamiokande \cite{Shiozawa:1998si,Nishino:2009gd} is performed with water
Cerenkov detectors in such a way that  8 out of 10 of the decaying protons are embedded in
an oxygen nucleus, where their spectrum is modified. In a pure shell model description corrections
are  applied to incorporate the shell model momentum distribution and corresponding binding of
the shell model orbits of the oxygen nucleus. However, Yamazaki and Akaishi \cite{YA99} pointed
out that this procedure does not take into account the correlations of
the decaying proton with the neighboring nucleons. Using a correlation function deduced from the
Reid soft core potential, they evaluated the effect on the invariant mass spectrum which acquires
a broad low energy tail representing $\simeq 10\%$ of the total decay.

It is customary in the decay problem to introduce the missing three momentum
$ P_{miss} = \Sigma_i P_i$,
sum of the momenta of the decay particles, and the missing mass,
$M_{miss}^2 = (\Sigma_i E_i)^2 -( \Sigma_i P_i)^2$. These quantities also refer to the
decaying proton.
They are related to the momentum and energy of the residual nucleus which can be in an excited
state defined by: $P_{miss} =P_{A-1}$ and $ \Sigma_i E_i =E^*_{A-1}$.  Since the state of the
residual nucleus
for each decay event is not known, there is no strict  constraint to identify a proton decay.
In the pure
 shell model case where the smearing is already present, the effect is rather mild and controllable.
But as shown in Ref. \cite{YA99} this is not the complete story. It is the aim of this work to evaluate
these distributions. The issue at stake is if the broadening is so big that a substantial fraction of
genuine decay event are lost in the background. In other words the lower limit on the proton lifetime
deduced  from the absence of events within a certain domain in the missing momentum and missing energy
variables must take into account the portion of  decay events outside this domain. It is therefore
useful to have the best possible probability distributions. Our result applies to any decay channel but
for illustrative purposes we will often refer to the pionic channel. It is valid as well for neutron
decay with disappearance of the hadron.

In  order to give a feeling for the importance  of the modification introduced by correlations, in
terms of particle-hole (ph) excitations, correlations translate into the existence of 2p2h excited
states mixed into the nuclear ground state. The decay of a correlated particle leaves the nucleus into
an excited state with one hole and one particle-one hole, for which we want to evaluate the excitation
energy. Beyond the  energy associated with one hole creation as evaluated in the shell model,  there is
the energy of the particle-hole which can be  approximated by $E_ {ph} = P_{ph}^2/2M$ where $P_{ph}$ is
the momentum exchanged between the correlated pair. Neglecting the momentum of the hole which has a
relatively narrow distribution, $P_{ph}$ is also the opposite of the missing momentum.

The missing mass square is then given by
\beqy
M_{miss}^2&=&(M^* - P^2_{miss}/(2M))^2 - P^2_{miss}\nonumber\\
\label{mmiss}&\simeq&M^{*2} - 2  P^2_{miss}
\eeqy
where $M^*$ is the nucleon mass reduced by the energy
necessary for the hole creation. The missing
mass square evolves approximately parabolically  with  $P_{miss}$.
To illustrate the expected effect let us  take an approximate value
$M^*\simeq 900$ MeV.  For a typical exchanged momentum  $P_{ph} =300$ MeV/c, which is also the
value of the missing momentum,  the missing mass value turns out to be $800$ MeV.  These two
missing values happen to be on the border line of the domain in which  Refs.
\cite{Shiozawa:1998si, Nishino:2009gd}
interpret an event as a proton decay one (no event in fact fell into this domain). We therefore
expect that in the analysis a substantial fraction of the correlated decays escapes detection.
Moreover, the future experiments aiming to improve the current limits on the proton decay
will have to introduce even tighter cuts to avoid the background due to the atmospheric  neutrino
interactions making the effects discussed even more important. The above qualitative argument is made
quantitative in the next section.
\section{Proton spectral function}

Since the decay of the bound proton occurs instantaneously on the scale of nuclear interactions one can
express the quantities relevant for the bound proton decay in terms of the nuclear spectral function
\beq
S_A(k,E) =\left|\left <\psi_A\left|\,a(k)
\,\delta(H_{A-1}- E)\,a^\dagger(k)\,\right|\psi_A\right>\right|^2
\eeq
which describes the probability to find a nucleon in the nucleus with momentum $k$ and produce a
residual $A-1$ system with excitation energy $E$ after an instantaneous removal of this nucleon.
The spectral function is related to the single nucleon momentum distribution as
\beq
\int dE\,S_A(k,E)\,=\,n_A(k),
\eeq
and it is normalized as
\begin{equation}
\int dE\,d\Vec{k}\,S_A(k,E)\,=\,1.
\end{equation}
In order to resolve the spectral function at the high resolution relevant for the proton decay one needs
to use probes which transfer  large energies and momenta, above 1 GeV, to the nucleons in the
nuclei. Such studies were performed in the last few years using proton and electron beams of
high energies.

It was observed \cite{egiyan} that the ratios of $(e,e^\prime)$ cross sections off nuclei and
the deuteron ($^3He$) are independent of $x$, $Q^2$ for $1.3 < x < 2$ and $Q^2 \ge 1.5$ $\mbox{GeV}^2$
corresponding to the kinematics where the electron can scatter only
off the correlated nucleon - nucleon pair with internal momenta $\ge 300$ MeV/c.
Moreover in $(e,e^\prime p)$ or $(p,2p)$ reactions on nuclei at large $Q^2$, a strong correlation
was observed between the emission of a fast proton and that of a nucleon (predominantly neutron)
in the opposite direction \cite{shneor, tang}. These studies confirmed theoretical
expectations of the presence of significant short-range correlations (SRC) in nuclei - for
instance in $^{12}C$ the probability $P_{^{12}C}$ to find a nucleon with momentum $\ge$
300 MeV/c is a factor of $\sim 5 \pm 0.5$ larger than in the deuteron.
The current models of the deuteron give  $P_D $ in the range $3 \div 4 \%$, and this corresponds
to $P_{^{16}O} = .15 \div .2$. The data also support the expectation that most of this probability
is due to the pn - tensor correlations (see e.g. \cite{Alvioli:2007zz, Schiavilla:2006xx}), which
are a specific case of 2p-2h excitations. For a review and detailed references see
\cite{Frankfurt:2008zv}.

In the  many-body models of nuclei with realistic NN potential the high momentum component
with momenta between 300 and 600 MeV/c originates from the interplay of attraction and
repulsion at distances $\le 1.2$ fm. Hence we have used two spectral functions \cite{ciofi,
Benhar:2005dj}  calculated in such models to analyze the effect of the nuclear
structure on the detection of the bound proton decay.

For the purposes of the analysis of the proton decay events it is convenient to choose as variables
the three-momentum of the decaying proton and the square of the bound proton mass $V=M^2_{miss}$
which fixes its off-shellness.
The two spectral functions of Ref. \cite{ciofi, Benhar:2005dj} have an uncorrelated
part,  $S_0(P_{miss},E)$, and a correlated one, $S_1(P_{miss},E)$, where $E$ is the proton removal
energy defined as $E=E_A-E^\star_{A-1}$. In the first model we use  $S_0$ as calculated with the
Skyrme force and renormalized by a factor 0.8.
The correlated part, $S_1$ represents 20\% of the total spectral function and it is given
by the model of \cite{ciofi}.
In this model $\int d^3 k dE S_1(k, E) \theta(k- k_0)= 11\%$ where $k_0=300$ MeV/c;
(the second model we considered provides the same result as far as this quantity is concerned).
The ratio of $^{16}O$ and deuteron high momentum components in these models varies in the
range of 3 $\div$ 6 for $300 < k  <  600 MeV/c$ which is rather close to the value of the
ratio $\sim 5$ obtained from the analysis of the hard phenomena and in particular $x > 1$
data (see the review in \cite{Frankfurt:2008zv}).
Smaller value of the total probability than in a phenomenological estimate is mainly due to
a later onset of the dominance of the short-range correlation regime.
The model of Ref. \cite{ciofi} for the correlated part of the spectral
function is based on the notion of the factorization of the two-body momentum distribution
for high values of the relative and small values of the center of mass momenta of the pair
and it is valid in this regions; this factorization was justified within a many-body approach
in Ref. \cite{Baldo:1900zz} and shown to hold for $^{16}O$ within the many-body calculation
of Ref. \cite{Alvioli:2007zz}.
It also gives a correct dependence for the center of mass of the correlated pair, as measured in
Refs. \cite{shneor} and \cite{tang}. We choose for the relative motion of the pair in the two-body
momentum distribution a parametrization which reproduces well the high momentum tail of
the deuteron in the region of interest and leads to a good description of the high momentum
tail of $n_{^{16}O}(k)$ \cite{Alvioli:2005cz, Pieper:1992gr}.

The spectral function of Ref. \cite{Benhar:2005dj} has been obtained within the Local Density
Approximation \cite{bffs}, in which the $(e,e^\prime p)$ data on single nucleon knock-out at
low missing energy \cite{saclay} is combined with
%------------------------------------------------------------------------------%
%--------------------------------------------------------------------------Fig1%
\begin{figure}[!ht]
\vskip -1.0cm
\centerline{\hspace{0.0cm}\includegraphics[width=9.0cm]{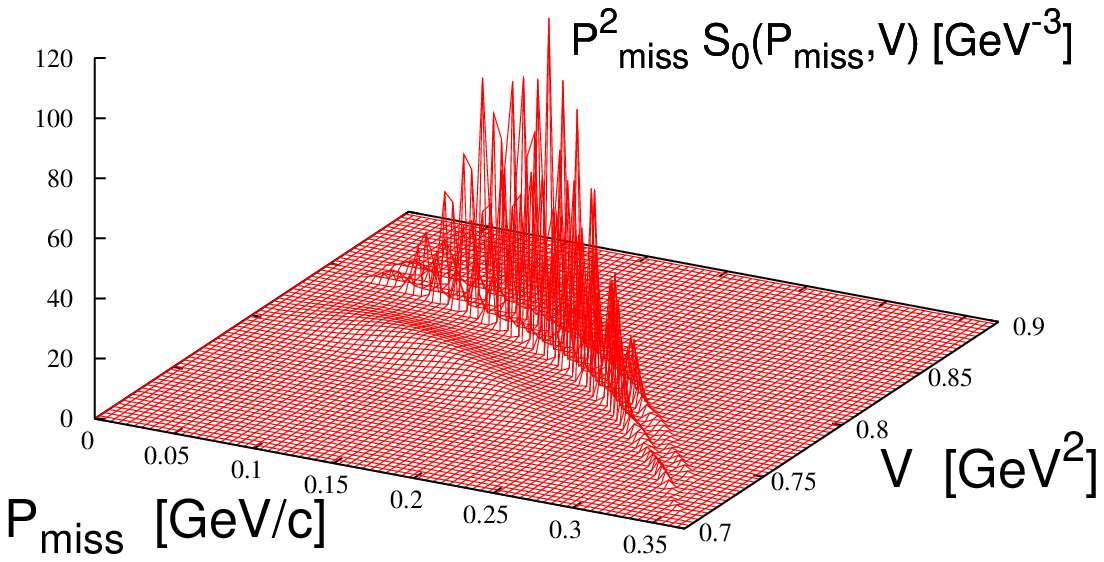}}
\vskip -1.0cm
\centerline{\hspace{0.3cm}\includegraphics[width=9.0cm]{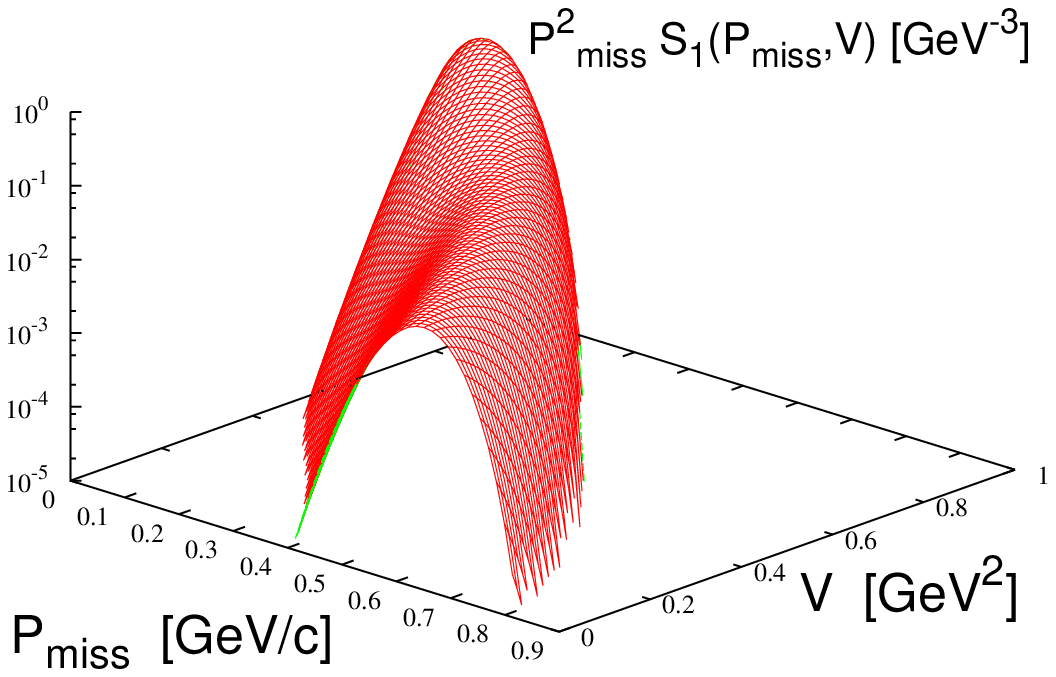}}
\caption{\textit{Top}: the \textit{uncorrelated} proton spectral
  function of $^{16}O$, $P^2_{miss}\,S_0(P_{miss},V)$, as calculated with
  the Skyrme force and plotted \textit{vs.} missing momentum $P_{miss}$ and
  virtuality, $V=M^2_{miss}$. \textit{Bottom}: the \textit{correlated} proton
  spectral function of $^{16}O$ from \cite{ciofi},
  $P^2_{miss} S_1(P_{miss},V)$.}\label{Fig01}
\end{figure}
%------------------------------------------------------------------------------%
%------------------------------------------------------------------------------%
the results of accurate theoretical calculations of the nuclear matter spectral function
at different densities \cite{bff}.
A direct measurement of the correlation component of the spectral function of $^{12}C$,
obtained measuring the $(e,e^\prime p)$ cross section at missing momentum and energy up to
$\sim$ 800 MeV and $\sim 200$ MeV, respectively, has been recently carried out at Jefferson Lab by the
E97-006 Collaboration \cite{E97-006}.
The data resulting from the preliminary analysis appear to be consistent with the theoretical
predictions based on the spectral function of Ref. \cite{Benhar:2005dj}.

\begin{figure}[!ht]
\vskip -1.0cm
\centerline{\hspace{0.3cm}\includegraphics[width=9.0cm]{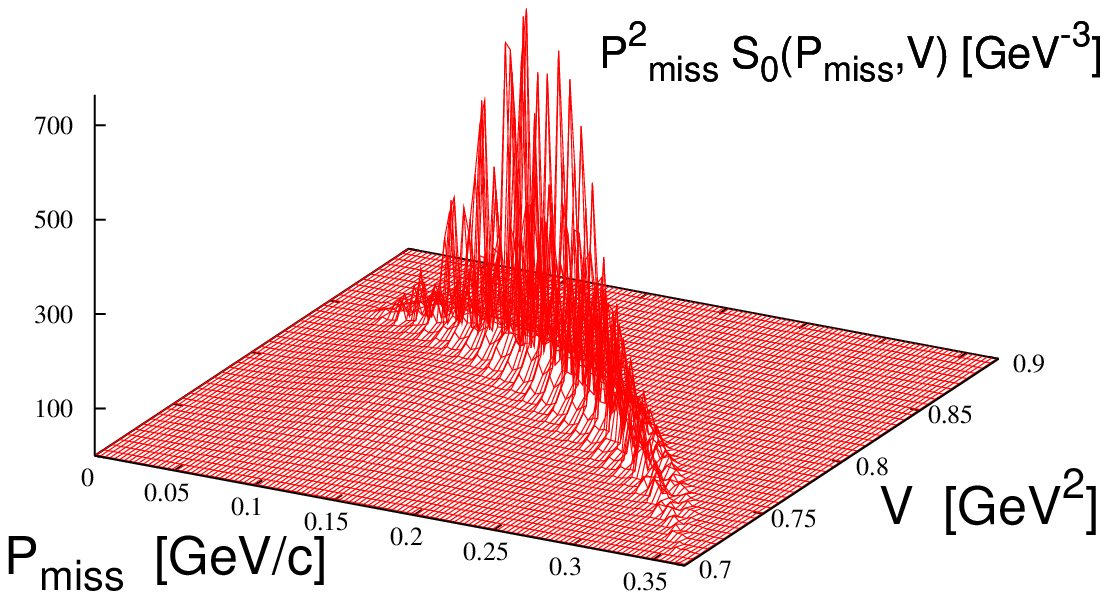}}
\vskip -1.0cm
\centerline{\hspace{0.3cm}\includegraphics[width=9.0cm]{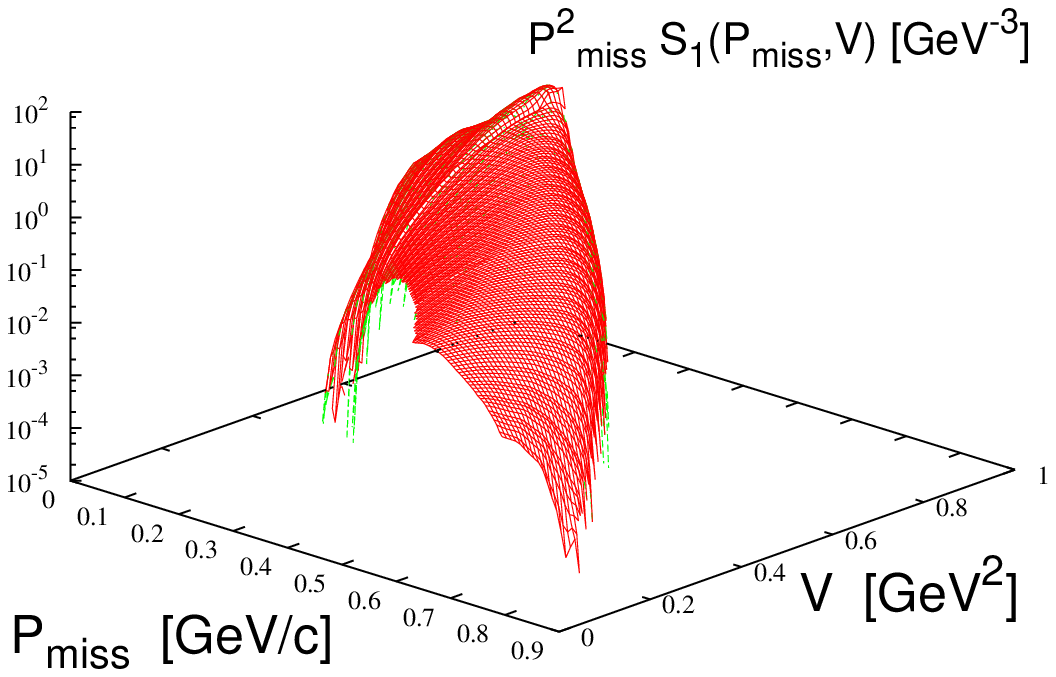}}
\caption{As in Fig. \ref{Fig01}, using the model of Ref. \cite{Benhar:2005dj}
  for both the uncorelated and correlated parts.}\label{Fig02}
\end{figure}

\begin{figure}[!ht]
\centerline{\includegraphics[width=9.5cm]{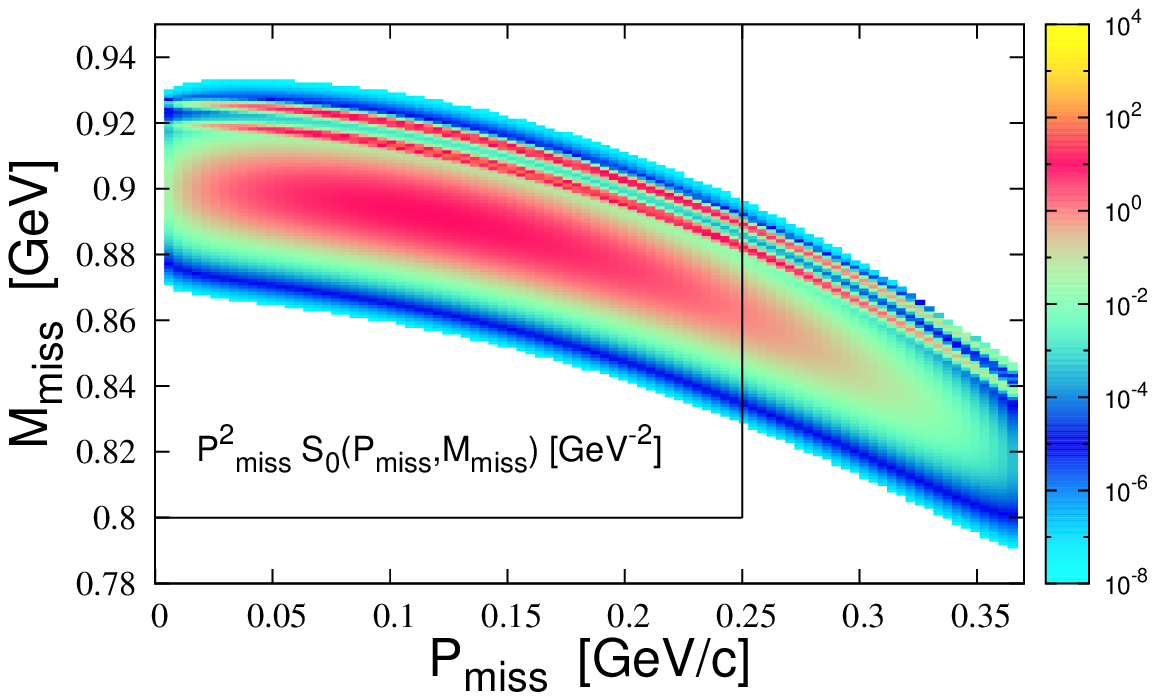}}
\vskip -1.0cm
\centerline{\includegraphics[width=9.5cm]{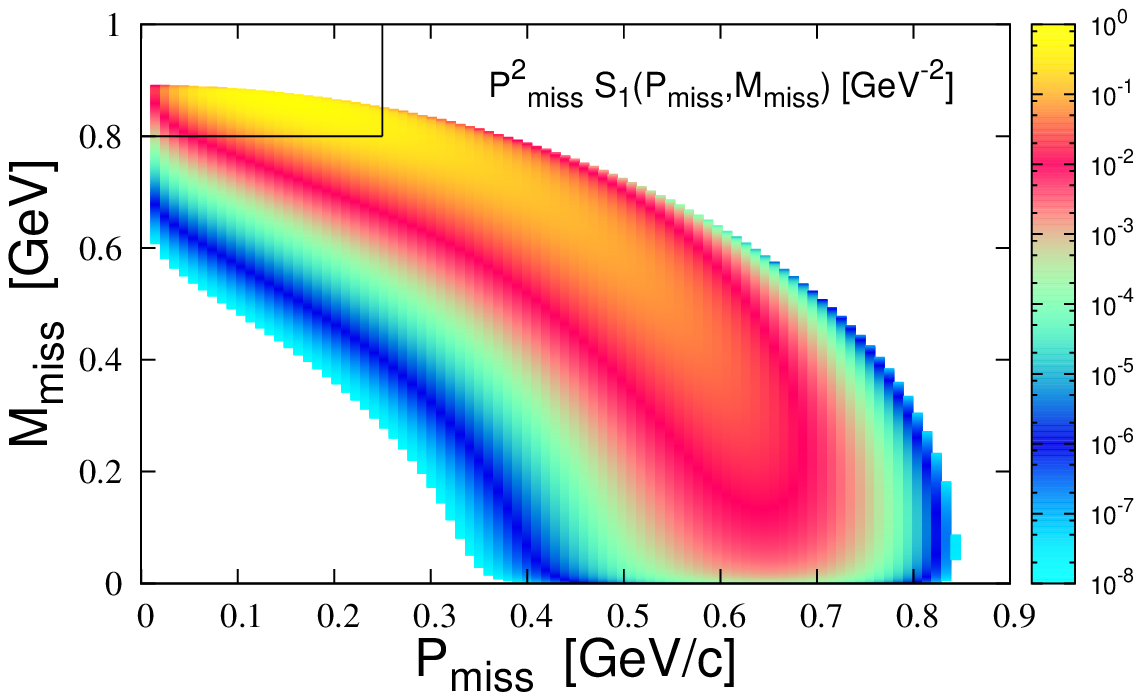}}
\caption{\textit{Top}: contour plot of the \textit{uncorrelated} proton spectral
  function of $^{16}O$, $P^2_{miss}\,S_0(P_{miss},M_{miss})$, as calculated with
  the Skyrme force and plotted \textit{vs.} missing momentum $P_{miss}$ and missing
  mass, $M_{miss}=\sqrt{V}$. \textit{Bottom}: the \textit{correlated} proton
  spectral function of $^{16}O$ from \cite{ciofi},
  $P^2_{miss} S_1(P_{miss},M_{miss})$.
  We show with black solid lines the cut $P_{miss} < 250$ MeV and $V < 640$ MeV,
  quoted in Refs. \cite{Shiozawa:1998si} and \cite{Nishino:2009gd} in both panels.}
\label{Fig1}
\end{figure}

The quantity $P^2_{miss}S_0(P_{miss}, V)$ is shown in  Figs. \ref{Fig01}, \ref{Fig02}
in three dimensional plots, and $P^2_{miss}S_0(P_{miss}, M_{miss})$ is shown
in Figs. \ref{Fig1}, \ref{Fig2} in contour plots, for the two considered models.
The strength is concentrated over three or four stripes in the $P_{miss}$, $M_{miss}$ plane in the
two cases; they correspond to the occupied shells of $^{16}O$: the $P_{1/2}$, $P_{3/2}$ and $S_{1/2}$
states in the case of calculation with the Skyrme force and to an additional $P_{3/2}$ state in the
case of Ref. \cite{Benhar:2005dj}.

The energies and widths of the occupied states in the two models we considered are the following.
In the first model, as calculated from the Hartree-Fock Skyrme model with shell model parameters
which describe the $(p,2p)$ and $(p,pn)$ data of Refs. \cite{Belostotsky:1987bb,Liyanage:2000bf},
they are $12.06$ MeV with a width $\simeq 5$ MeV for the $P_{1/2}$ state,
%------------------------------------------------------------------------------%
%--------------------------------------------------------------------------Fig1%
%------------------------------------------------------------------------------%
%------------------------------------------------------------------------------%
the  $P_{3/2}$ state has $18.63$ MeV with a width of $5$ MeV; the $S$ state
is quite broad with a width of $\simeq 40$ MeV for an energy $37.96$ MeV.
In the second model, Ref. \cite{Benhar:2005dj}, we have the following energies, with roughly
the same widths:
$P_{1/2}$ at $12.50$ MeV, $P_{3/2}$ at $18.75$ MeV, a second $P_{3/2}$ at $23.00$ MeV and $S_{1/2}$
at $42.50$ MeV.
The model takes into account a rather small contribution of higher energy excitations
where intermediate states with one particle in the continuum and one hole in the final state.

The behavior in the momentum $P_{miss}$ can be inferred from the expression of $V$ which
in the uncorrelated case is:
\beqy
V&\equiv&M^2_{miss}\,=\,(P_{\mu\,A}\,-\,P_{\mu\,A-1})^2\nonumber\\
\label{virt}&=&\left(M_p\,-\,E\,-\,\frac{P^2_{miss}}{2 M_{A-1}}\right)^2
\,-\,P^2_{miss}\,.
\eeqy
Expanding the square in Eq. (\ref{virt}) and neglecting the $P^4_{miss}$ term we obtain:
\beqy
V(P_{miss})&\simeq&(M_p\,-\,E)^2\,-\,
\,\frac{M_p\,-\,E\,+\,M_{A-1}}{M_{A-1}}\,P^2_{miss}\nonumber\\
&=&(M_p\,-\,\epsilon_\alpha)^2\,-\,C_\alpha\,P^2_{miss},
\label{fit0b}
\eeqy
where  $\epsilon_\alpha$ are the values of the proton shells energies.
The coefficients $C_\alpha$ are $1.07$ and $1.06$  for the $P$ and $S$ proton shells, respectively.
%------------------------------------------------------------------------------%
%--------------------------------------------------------------------------Fig1%
%------------------------------------------------------------------------------%
%------------------------------------------------------------------------------%
As for the correlated spectral function, $P^2_{miss}S_1(p_{miss}, V)$ is represented
in Figs. \ref{Fig01}, \ref{Fig02} in three dimensional plots, and
$P^2_{miss}S_1(p_{miss}, M_{miss})$ is shown in Figs. \ref{Fig1}, \ref{Fig2} in
contour plots; it has a similar behavior than the corresponding uncorrelated
quantity but it is broader.
One can see from Figs. \ref{Fig1}, \ref{Fig2} that the correlated spectral functions in the two
models exhibit differences; a detailed comparison of the two spectral functions
is out of the scope of the present paper. Nevertheless, these differences
do not affect our conclusions, as appears in the following, since most of the strength
is concentrated along the stripe of maximum strength, even if the second model is more peaked at
low momenta and it is more narrow around the center of the stripe. The center of the corresponding
stripe obeys the following equation in the $V, P_{miss}$ plane for both models:
%------------------------------------------------------------------------------%
%--------------------------------------------------------------------------Fig2%
\begin{figure}[!ht]
\centerline{\includegraphics[width=9.5cm]{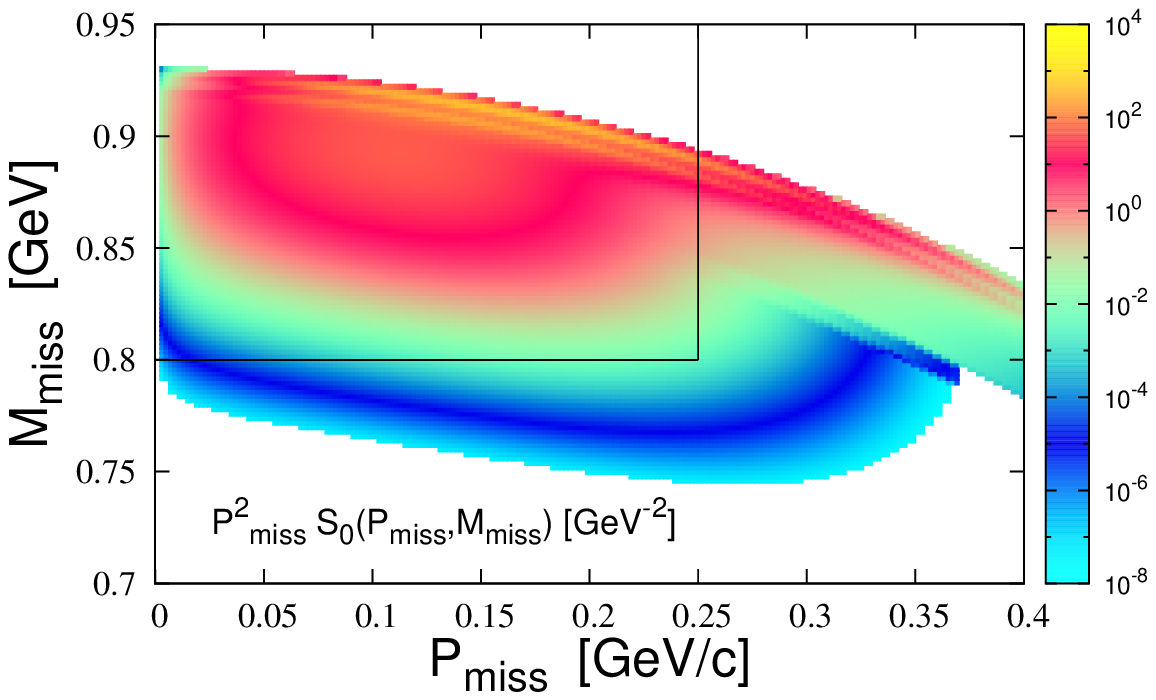}}
\vskip -1.0cm
\centerline{\includegraphics[width=9.5cm]{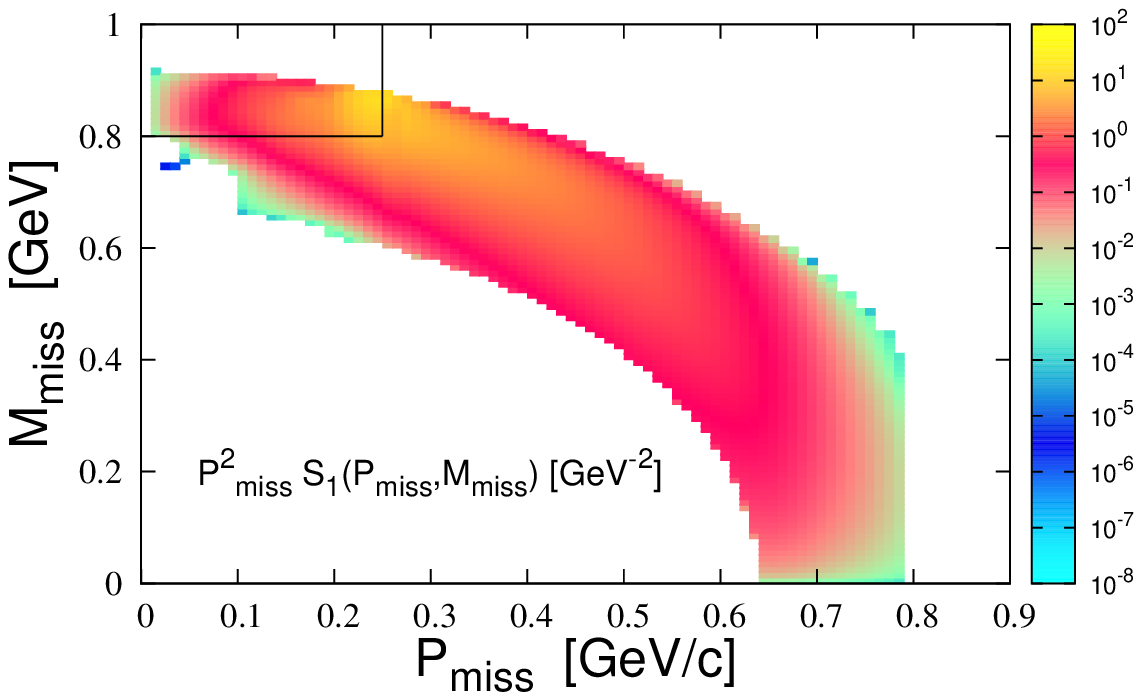}}
\caption{As in Fig. \ref{Fig1}, but with the model spectral function of
    Ref. \cite{Benhar:2005dj}.}
\label{Fig2}
\end{figure}
%------------------------------------------------------------------------------%
%------------------------------------------------------------------------------%
\beqy
\label{fit1}V(P_{miss})&=&0.78\,-\,1.78\,P^2_{miss}
\eeqy
where $V$ is expressed in $\mbox{GeV}^2$ and $P_{miss}$ in GeV/c, which is close to the
approximate expression of Eq. (\ref{mmiss}).

The concentration of the strength of the spectral function in limited regions of space which
project in some bands in the $M_{miss}$, $P_{miss}$ plane suggests a complementary analysis of
the data specifically aimed at the decay of the $^{16} O$ protons. It consists in the following:
to look for events which, in this plane, fall in one or several, depending of the accuracy of the
data, regions of this plane selected to cover the lines of the maximum of the (uncorrelated or
correlated) strengths defined in Eqs. (\ref{fit0b}, \ref{fit1}) so as to maximize the number of
significant events while minimizing the background, \textit{i.e.}, the area. Correlated events can
also be included in this way.
The calculated proton spectral function in $^{16}O$ is sufficiently reliable, as it is established in
connection with various nuclear physics experiments, to allow for this possibility.

This kind of analysis, if feasible, precludes the subsequent distortions of the pion kinematics after
emission by the proton. We will comment later on that.
%%------------------------------------------------------------------------------%
%%--------------------------------------------------------------------------Fig3%
\begin{figure}[!ht]
\centerline{\includegraphics[width=9.5cm]{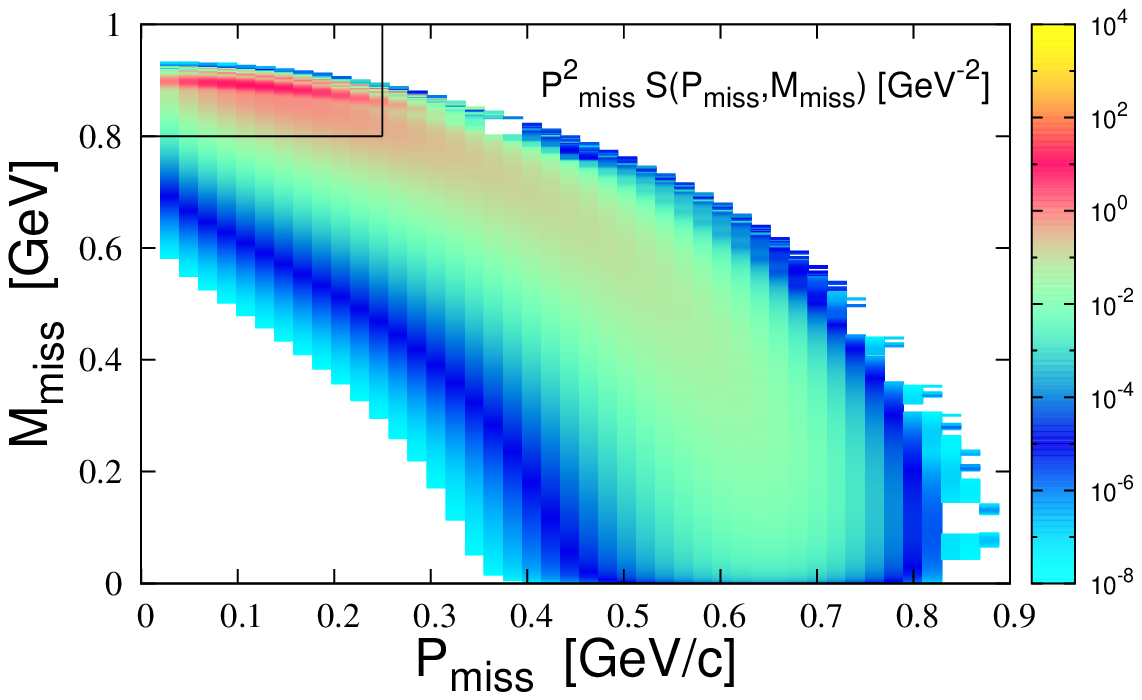}}
\vskip -1.0cm
\centerline{\includegraphics[width=9.5cm]{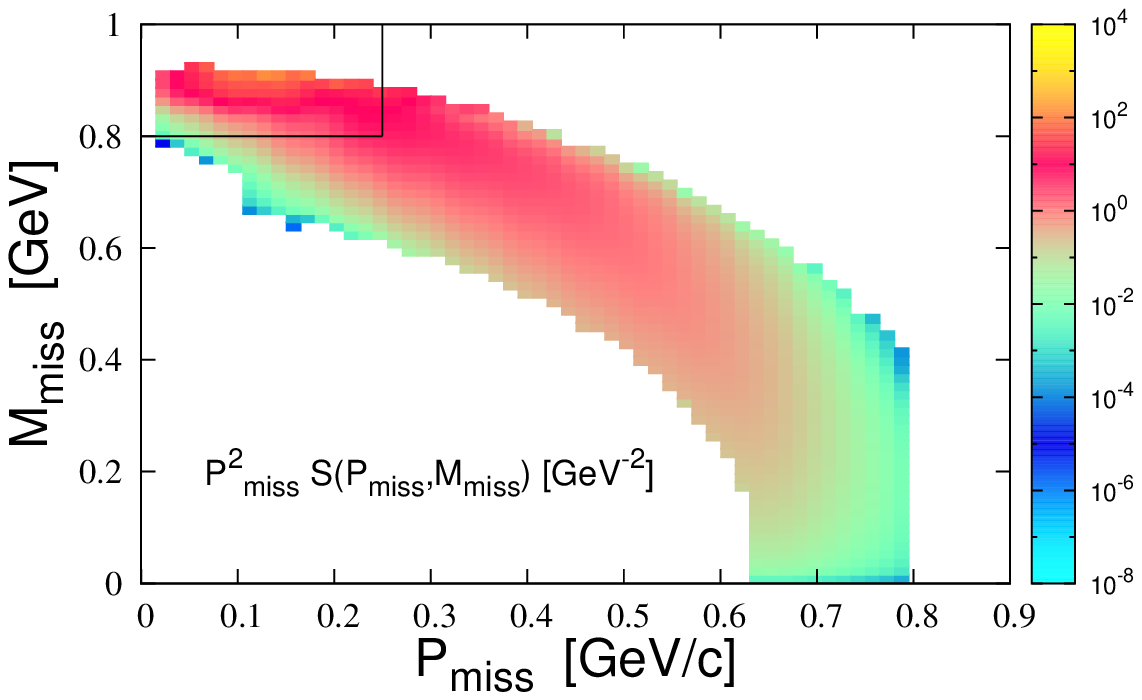}}
\caption{The total spectral function, $S_0+S_1$, within the two considered models.
  \textit{Top}: the mixed model of $S_0$ from the Skyrme force and $S_1$ from Ref.
  \cite{ciofi};
  \textit{bottom}: the model of Ref. \cite{Benhar:2005dj}.
  }
\label{Fig3}
\end{figure}
%------------------------------------------------------------------------------%
%------------------------------------------------------------------------------%
A technical remark: for the correlated part, when transforming coordinates from $(P_{miss},E)$ to
$(P_{miss},V)$ as in Eq. (\ref{virt}), we impose that $V$ stays positive, so we forbid a certain
region of $E,P_{miss}$ space to be accessible; this in turn means the normalization of the correlated
spectral function is not exactly $0.2=\int dE d\Vec{P}_{miss}\,S_1(P_{miss},E)$ but
\beq
\int dV d\Vec{P}_{miss}\,S_1(P_{miss},V)\,=\,0.18\,.
\eeq
We have checked that integrating over negative values of $V$ gives the missing normalization,
$\int^0_{-\infty} dV \int d\Vec{P}_{miss}\,S_1(P_{miss},V)=0.02$.

\begin{figure}[!ht]
\centerline{\hspace{0.3cm}\includegraphics[width=9.5cm]{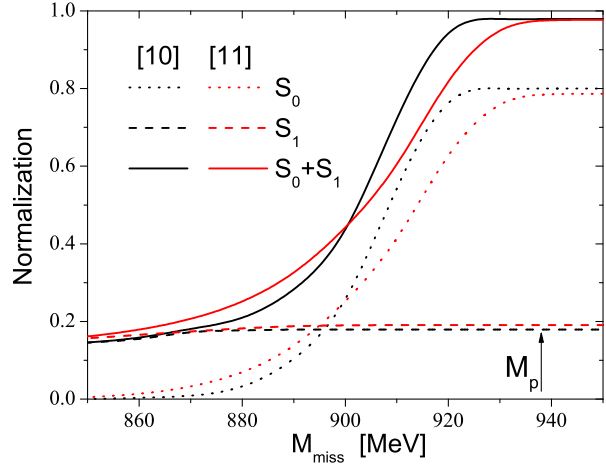}}
\caption{The normalization of the spectral function after the
  cut of Eq. (\ref{mmiss}), as a function of the upper limit of integration
  in $(M_{miss})_{max}=\sqrt{V_{max}}$; the arrow shows the position of the proton
  mass, $M_p$. Black curves correspond to the model of Ref. \cite{ciofi} and
  the red curves correspond to the model of Ref. \cite{Benhar:2005dj}; separate 
  contributions are shown for the uncorrelated/correlated part within both models.}
\label{Fig4}
\end{figure}

In Fig. \ref{Fig3} we show the total spectral function $S_0 +S_1$, in the two considered
models, as a function of $P_{miss}$ and $M_{miss}$, while
in Fig. \ref{Fig4} we present the normalization integral of the spectral function
\beq
\label{normcut}
N(V_{max})\,=\,\int^{V_{max}}_0 dV \int d\Vec{P}_{miss}\,S(P_{miss},V).
\eeq
It represents the number of events lost by applying a cut on the missing mass such that only
the events which correspond to a missing mass larger than this particular value $M_{miss}$
are kept, irrespective of the momentum.
%------------------------------------------------------------------------------%
%--------------------------------------------------------------------------Fig4%
%------------------------------------------------------------------------------%
%------------------------------------------------------------------------------%
If no cut on $V$ is applied $N=0.98$. If we use the cut $\sqrt{V}\ge 800$ MeV, $k\le 250$ MeV/c,
presented in Figs. \ref{Fig1}, \ref{Fig2} and \ref{Fig3}, the number of nuclear events is
reduced by a factor of $N=0.83$ using the model of Ref. \cite{ciofi} and by a factor
of $N=0.80$ using the model of Ref. \cite{Benhar:2005dj}.
However, future experiments are likely to have to introduce tighter cuts in order to reduce
the background
from the interactions of atmospheric  neutrinos. If, for example,  the cut $\sqrt{V} > 900$ MeV is
imposed, $\simeq 44\%$ of the events are removed, namely $26\%$ ($25\%$) from uncorrelated events
and $18\%$ ($19\%$), the near totality of correlated events in the first (second) considered model.
With a tight constraint on $M_{miss}$ the fraction lost is quite appreciable if no other precaution
is taken. The correlated analysis that we discussed may allow a better efficiency.

There are other effects which reduce the contribution of the bound nucleon decays.
This includes a reduced phase volume which is $\propto \sqrt{V}$ for decays with production
of light particles. Furthermore the very  mechanism of the proton decay may be sensitive to the
nuclear correlations. For example, if the decay amplitude is proportional to the three quark
wave function at the origin, see \textit{e.g.} \cite{Berezinsky:1981qb}, the effect of suppression of
the point-like configurations in bound  nucleons \cite{Frankfurt:1985cv,Frank:1995pv} would
contribute, reducing the rate of the decay by about 14\% for the  $M_p - \sqrt{V} =100$ MeV
cut.

All the effects that we have discussed are genuine medium effects on the decay amplitude.
They are not the whole story. The subsequent history of the  pion, rescatterings / absorption
in oxygen further reduces the number of ``observable'' pions. The inelastic scattering of pions
clouds the message on the kinematics since the inelastically scattered pion  ejects a nucleon.
The corresponding point in the $M_{miss}$, $P_{miss}$ plane would be likely to fall outside the
interesting regions delimited from the proton spectral function that we discussed in this work.
Therefore for an analysis of the type suggested in this work, inelastically scattered pions
may be considered as lost events. It may represent a reduction factor of about $0.6$.
This is usually taken care of through a Monte Carlo evaluation which is beyond the scope of
the present work.

In conclusion we have introduced, in the problem of the identification of the decay events of
the protons bound in the oxygen nucleus, the use of proton spectral function. 
It allows the prediction
of the location in  the  $M_{miss}$, $P_{miss}$ plane of the decay events. Our spectral function
has an uncorrelated and a correlated part. It is has been tested against a number of nuclear physics
experiments and the reliability of our prediction is sufficient to be exploitable. We considered
two models for the spectral function and the conclusions on the correlated cuts holds even if the
two models exhibit some difference for the spectral functions.
It appears that for the future nucleon decay experiments with tight cuts on the mass of the products
of the proton decay it may be interesting to consider, as a complementary information for the decay
events of the oxygen protons, correlated cuts on the mass and missing momentum obtained from the
spectral function in order to decrease  the background to signal ratio. The price to pay for this
type of analysis is the loss of decay events where a pion produced in the decay is inelastically
scattered, which clouds the reconstitution of the proton spectral function in oxygen. The loss in
intensity however is moderate and it may be compensated by the advantage of a decrease of the
background.\\

We thank G. Chanfray, C. Ciofi degli Atti, D. Davesne, M. Fidecaro, L. Frankfurt,
M. Martini and M. Zhalov for useful discussions.

%------------------------------------------------------------------------------%
%------------------------------------------------------------------------------%

%------------------------------------------------------------------------------%
%------------------------------------------------------------------------------%
\end{document}